\documentclass[structabstract]{aa}
\usepackage{graphicx}
\usepackage{txfonts}
\usepackage{longtable}
\usepackage{natbib}
\bibpunct{(}{)}{;}{a}{}{,}

\begin{document}

\title{A distant OB association around RAFGL 5475\thanks{Based on observations obtained at the Centro Astron\'omico Hispano-Alem\'an, Calar Alto, Spain}
}
\author{F. Comer\'on\inst{1}
\and A.A. Djupvik\inst{2}
\and J. Torra\inst{3}
\and N. Schneider\inst{4,5}
\and A. Pasquali\inst{6}
}
 \institute{
  European Southern Observatory, Alonso de C\'ordova 3107, Vitacura, Santiago, Chile\\
  \email{fcomeron@eso.org}
  \and
  Nordic Optical Telescope, Apdo 474, E-38700 Santa Cruz de La Palma, Spain
  \and
  Institut de Ci\`encies del Cosmos (ICCUB-IEEC), Barcelona, E-08028, Spain
  \and
  I. Physik Institut University of Cologne, D-50937 Cologne, Germany
  \and
  OASU/LAB-UMR5804, CNRS, Universit\'e Bordeaux 1, F-33270 Floirac, France
  \and
  Astronomisches Rechen-Institut, Zentrum f\"ur Astronomie der Universit\"at Heidelberg, M\"onchhofstr. 12-14, D-69120 Heidelberg, Germany
  }
%
%
\date{Received; accepted}
\abstract
{Observations of the galactic disk at mid-infrared and longer wavelengths reveal a wealth of structures indicating the existence of complexes of recent massive star formation. However, little or nothing is known about the stellar component of those complexes.}
{We have carried out observations aiming at the identification of early-type stars in the direction of the bright infrared source RAFGL~5475, around which several interstellar medium structures usually associated with the presence of massive stars have been identified. Our observations have the potential of revealing the suspected but thus far unknown stellar component of the region around RAFGL~5475.}
{We have carried out near-infrared imaging observations ($JHK_S$ bands) designed to reveal the presence of early-type stars based on their positions in color-color and color-magnitude diagrams centered on the location of RAFGL~5475. We took into account the possibility that candidates found might belong to a foreground population physically related either to M16 or M17, two giant HII regions lying midway between the Sun and RAFGL~5475.}
{The near-infrared color-color diagram shows clear evidence for the presence of a moderately obscured population of early-type stars in the region imaged. By studying the distribution of extinction in their direction and basic characteristics of the interstellar medium we show that these new early-type stars are most likely associated with RAFGL~5475.
}
{By investigating the possible existence of massive early-type stars in the direction of RAFGL~5475 we have discovered the existence of a new OB association. A very preliminary assessment of its contents suggests the presence of several O-type stars, some of them likely to be associated with structures in the interstellar medium. The new association is located at 4 kpc from the Sun in the Scutum-Centaurus arm.}



\maketitle

\section{Introduction} \label{intro}

Despite the dramatic advances in the knowledge of our galactic neighborhood made possible in the past few decades by largescale surveys at all wavelengths, obscuration by dust continues to be a serious hindrance at the time of recognizing and characterizing some of the major stellar structures beyond several hundreds of parsecs from the Sun. Even Gaia, with its superb astrometric capabilities that hold the potential of revolutionizing our understanding of the Galaxy, is limited along the lines of sight near the galactic plane by the visible wavelength range in which it operates. All-sky or wide-area surveys in the near-infrared such as 2MASS \citep{Skrutskie06}, WISE \citep{Wright10}, or VVV \citep{Minniti10} are invaluable tools to probe within and beyond heavily obscured regions, but their limited sensitivity or the very limited information on the spectral energy distribution that they yield about most classes of stellar objects often constrain what can be learned about the stellar content of the solar neighborhood, and our knowledge of some interesting regions within a few kiloparsecs from the Sun remains scanty.

One of those regions is that around the bright infrared source RAFGL~5475, which remains virtually unexplored thus far. RAFGL~5475 is a compact HII region without an ultracompact component associated \citep{Codella95}, with maser emission in methanol \citep[6.7~GHz;][]{Szymczak00,Pestalozzi05} and water \citep[22.2~GHz;][]{Codella95}. Nothing of this region is seen at visible wavelengths due to strong obscuration by dust. Wide surveys of the galactic disk covering the area at various wavelengths list dense cores \citep{Schlingman11,Shirley13}, dark clouds \citep{Peretto09}, HI shells \citep{Furst90}, and HII regions \citep{Lockman89,Anderson11}. Despite the crowdedness along the line of sight in that general direction of the galactic disk, the narrow range of radial velocities (+46 to +50~km~s$^{-1}$) of the various tracers for which radial velocities can be measured indicate the existence of a coherent structure at the distance of 3.8-4.0~kpc \citep{Anderson09}, tracing the Scutum-Centaurus arm and kinematically well separated from the foreground structures associated with M16 and M17.

In Spitzer images at 5.8~$\mu$m and beyond, the field around RAFGL~5475 appears as a complex interplay of extended emission, infrared dark clouds silhouetted against it, large-scale filaments, and at least two clearly outlined bubbles. Herschel images at 250~$\mu$m reveal vast amounts of cold gas extending over a similar area but also beyond. All this points to the existence of a large-scale star formation region around RAFGL~5475, where we may expect the presence of massive stars interacting with their surrounding interstellar medium. We have adopted the kinematic distance of $4.0$~kpc (distance modulus 13.0~mag), similar to other star-forming regions in the area for which trigonometric distances have been derived through VLBI observations of associated masers \citep{Reid14}. The extent of the area covered by the structures related to RAFGL~5475, $\sim 20'$ across, translates into a physical size of $\sim 25$~pc, similar to the Orion Molecular Cloud complex. The structures present in the region suggest the existence of evolved groups of massive stars that have been able to already blow up bubbles reaching 5.2~pc across like C16.31-0.16 \citep{Anderson09}, coexisting with an ongoing massive star forming site like RAFGL~5475 itself, and potential sites of future star formation in the infrared dark clouds. Again, this bears clear similarities with the Orion Molecular Cloud, where different age groups of the Orion OB1 association are identified, and where we are witnessing the birth of a massive star in the Becklin-Neugebauer (BN) object \citep[e.g.,][]{Tan04}.

In this study we present $JHK_S$ ($1.2-2.3$~$\mu$m) observations of an area centered on RAFGL~5475 ($l = 16.36^\circ$, $b = -0.21^\circ$) looking for the possible evidence of hot, massive stars that may reveal the thus far unknown stellar component of RAFGL~5475. We do identify such a population and discuss some of its properties showing that it is most likely composed of the members of an extended OB association physically related to RAFGL~5475.

\section{Observations and data reduction} \label{obs}

The observations reported here were obtained using the PANIC wide field near infrared camera at the Calar Alto 2.2m telescope \citep{Baumeister08} on four nights between the fourth-to-fifth and seventh-to-eight August 2016. The camera has a detector mosaic of $2 \times 2$ detectors of $2048 \times 2048$~pixels$^2$ with a pixel scale of $0''450$ per pixel. However, the very large number of bad pixels of three of the detectors at the time of our observations made only one of the detectors useful, still yielding a considerable field of view of $15' \times 15'$ for our scientific purposes. We obtained a number of images through the $J$, $H$, and $K_S$ broadband filters.
Each image was built from the combination of frames on a dithered pattern of nine telescope pointings on a $3 \times 3$ square grid of $20'' \times 20''$, each of which consisted in turn of a stack of NDIT individual exposures of DIT seconds of integration time. Table~\ref{exposures} gives the parameters of the exposures composing each image.

\begin{table}[t]
\caption{Exposure parameters}
\begin{tabular}{cccccc}
\hline
filter & \# of  & dithered  & NDIT & DIT & total    \\
       & images & pointings &      & (s) & exp. time\\
\hline
$J$        & 144 & 9 & 12 & 5 & 8640 \\
$H$        & 144 & 9 & 6 & 10 & 8640 \\
$K_S$      & 45 & 9 & 6 & 10 & 2700 \\
\hline
\end{tabular}
\label{exposures}
\end{table}

Dome flat fields with the dome illumination on and off were used to correct for pixel-to-pixel sensitivity differences, and sequences of images obtained with the closed dome and a constant illumination but systematically increasing exposure times were used to calibrate the non-linearity of each detector pixel. For the latter, the pixel counts as a function of exposure time were fitted by a second-degree polynomial, which was found to provide a sufficiently accurate description of the pixel response.

Field distortion effects producing a slight change in the pixel scale over the field of view are clearly noticeable in our observations, thus preventing a direct shift-and-add stacking of the frames obtained at the different points of the dither pattern. To overcome this, a distortion map was produced by taking as a reference the cataloged positions of UCAC3 stars \citep{Zacharias10} identified in our images and using the GEOMAP and GEOTRANS tasks on IRAF to apply a flux-conserving distortion correction to all our frames. The distortion map was found to remain constant for all filters and nights. This distortion map was then applied to the flat-fielded, linearity-corrected images and the distortion-corrected images were used from that point on. Sky frames were produced by stacking with median filtering the images taken within each dithering sequence without correcting for the telescope offsets in between and were subtracted from the individual images produced at each point of the dither pattern. The sky-subtracted images were then stacked, this time applying a shift to correct for the telescope offsets between exposures. Finally, all the images obtained in the same filter were combined into a single deep image per filter, with the total exposure times indicated in Table~\ref{exposures}.

The deep images thus obtained in all filters were combined together into a single one, and stars were detected in it using the DAOFIND \citep{Stetson87} task layered on IRAF, thus obtaining a catalog of all the stars detected in the combined image. Point-spread function instrumental photometry using tasks in the DAOPHOT package was then obtained for each star of the catalog in the deep image for each of the five filters used. We then determined the transformation from the instrumental to the 2MASS photometric system using the photometry of stars from the 2MASS all-sky point source catalog appearing in the field of view of our images. The following transformation expressions were derived:


$$J = j + C_j + 0.0693 (J- H) -0.0177 (J-H)^2 \eqno{(1a)},$$
$$H = h + C_h + 0.0086 (H-K_S) -0.0313 (H-K_S)^2 \eqno{(1b)},$$
$${K_S} = {k_S} + C_k + 0.0078 (H-K_S) -0.0076 (H-K_S)^2 \eqno{(1c)},$$

\noindent where capitals refer to magnitudes in the 2MASS system, lower case are instrumental magnitudes, and $C_j$, $C_h$, $C_k$ are zeropoints in each filter. Our exposure parameters resulted in saturation for stars with magnitudes $J \lesssim 9.0$, $H \lesssim 9.5$, $K_S \lesssim 8.0$. For stars brighter than those limits, which are much brighter than the detection limits of the 2MASS catalog, the 2MASS magnitudes were directly adopted.

\begin{figure}[ht]
\begin{center}
\hspace{-0.5cm}
\includegraphics [width=9cm, angle={0}]{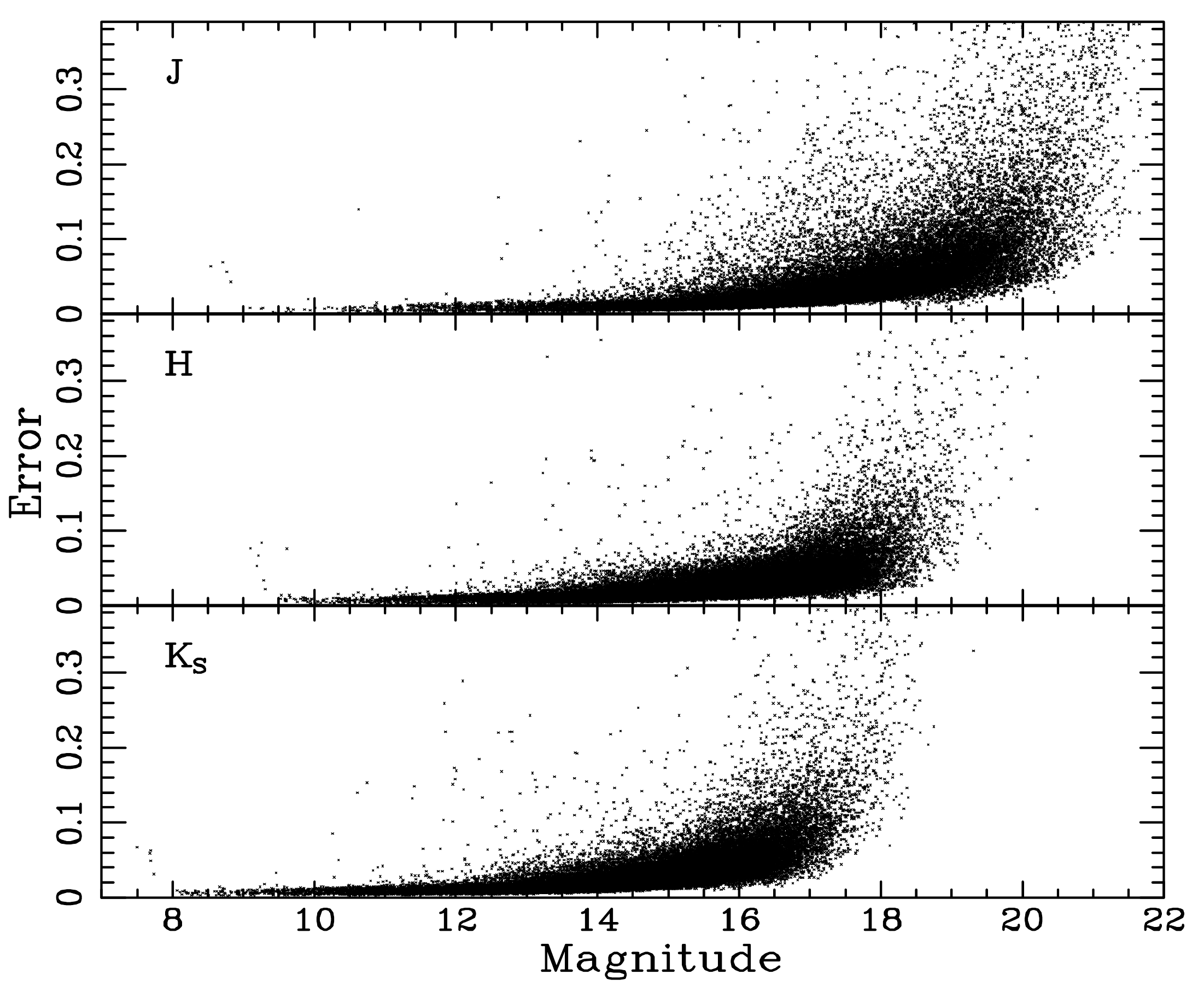}
\caption []{Individual errors as a function of the magnitude in each of the broadband filters. These are internal errors of the instrumental photometry, and do not include possible contributions from errors in the zeropoints or in the color transformation coefficients.}
\label{errors}
\end{center}
\end{figure}

The $5\sigma$ detection limits are estimated to be $J \simeq 19.5$, $H \simeq 18.0$, $K_S \simeq 16.5$. The errors as a function of the magnitude in each band are plotted in Figure~\ref{errors}. It may be noted that the stellar images in all our observations were affected by a moderate amount of coma aberration due to a slight misalignment of the primary mirror of the telescope with respect to the optical axis. This, together with the seeing during our observations and the high airmass through which the field had to be observed due to its Southern declination, resulted in a point-spread function in the combined images having a width of $2''2$ ($J$), $2''0$ ($H$) and $2''1$ ($K_S$). The combination of a non-optimal image quality and the crowdedness of the field limited the quality of the sky frames that could be constructed as described above and decreased the magnitudes where source confusion starts being noticeable, thus compromising the quality of the photometry at faint magnitudes. Nevertheless, a comparison of our color-magnitude and color-color diagrams with those produced from the 2MASS Point Source Catalog shows that our photometry is significantly better than those of 2MASS at a given magnitude, and the depth is significantly improved (see Sect.~\ref{results}). This improved quality turns out to be critical in enabling the analysis presented here.

\begin{figure*}[ht]
\begin{center}
\hspace{-0.5cm}
\includegraphics [width=12cm, angle={0}]{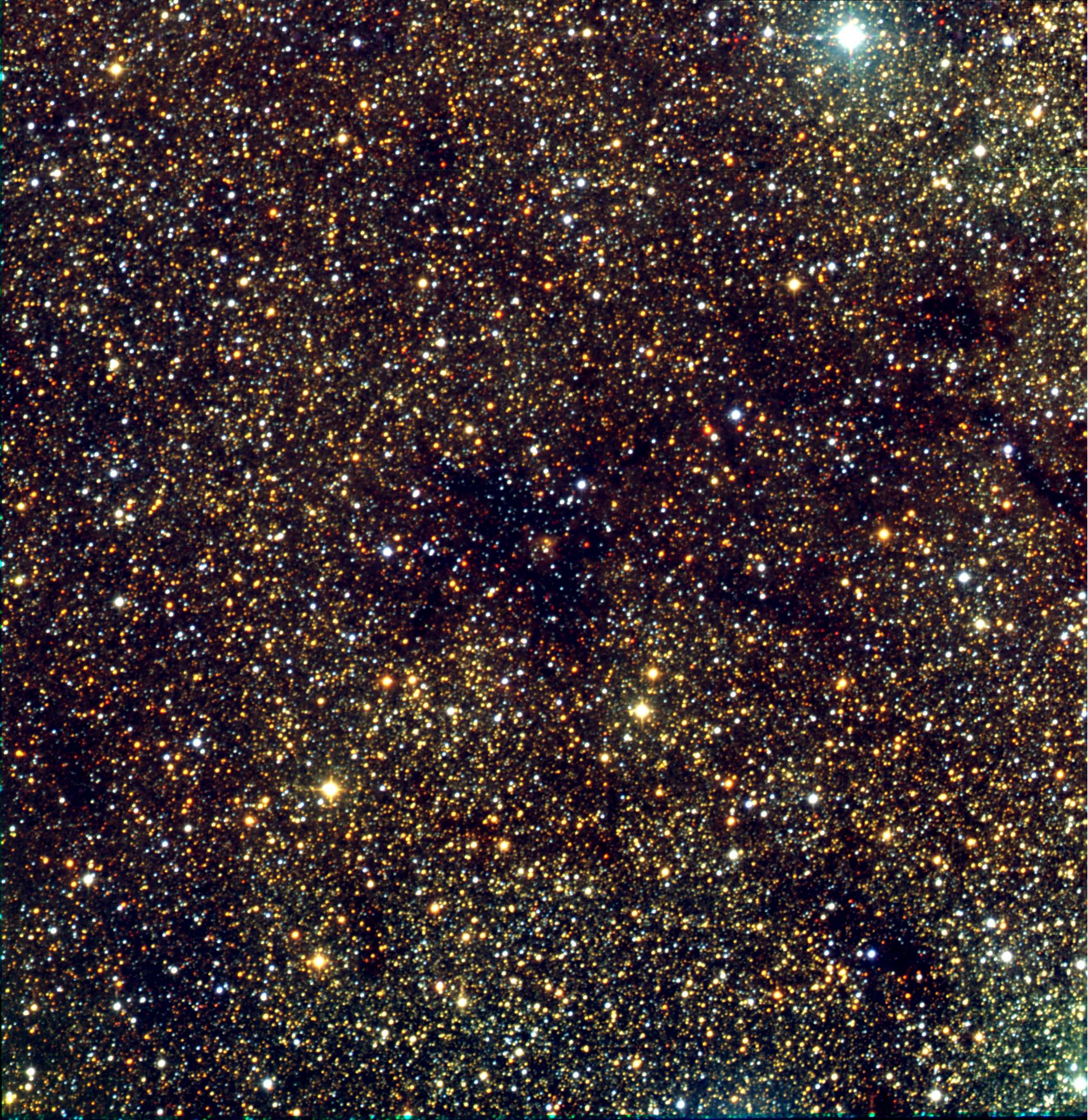}
\caption []{Color composite of the field of view covered by our near-infrared observations, with all the exposures obtained in a given filter stacked. The blue channel corresponds to the image taken through the $J$ filter, green through the $H$ filter, and red through the $K_S$ filter. The field is $15' \times 15'$, with north at the top and east to the left. RAFGL~5475 is a faint, slightly extended object near the middle of the dark cloud at the center of the image.}
\label{JHKs}
\end{center}
\end{figure*}

\section{Results\label{results}}

  Figure~\ref{JHKs} shows a color composite of the entire observed field combining all the images taken through the
$J$, $H$ and $K_S$ filters. RAFGL~5475 is visible in the center of the field and the dark cloud hosting it is also clearly delineated against the densely crowded background. No bright nebulosity is seen either in our broadband or narrow-band images other than the very compact one associated with RAFGL~5475, despite the existence of several HII regions in the area revealed by radio surveys \citep{Lockman89} and the abundance of nebulosity at wavelengths longer than 5~$\mu$m revealed by Spitzer images (see Sect.~\ref{discussion}), showing that those structures in the interstellar medium are considerably obscured by intervening dust even in the near-infrared.

\begin{figure}[ht]
\begin{center}
\hspace{-0.5cm}
\includegraphics [width=8.5cm, angle={0}]{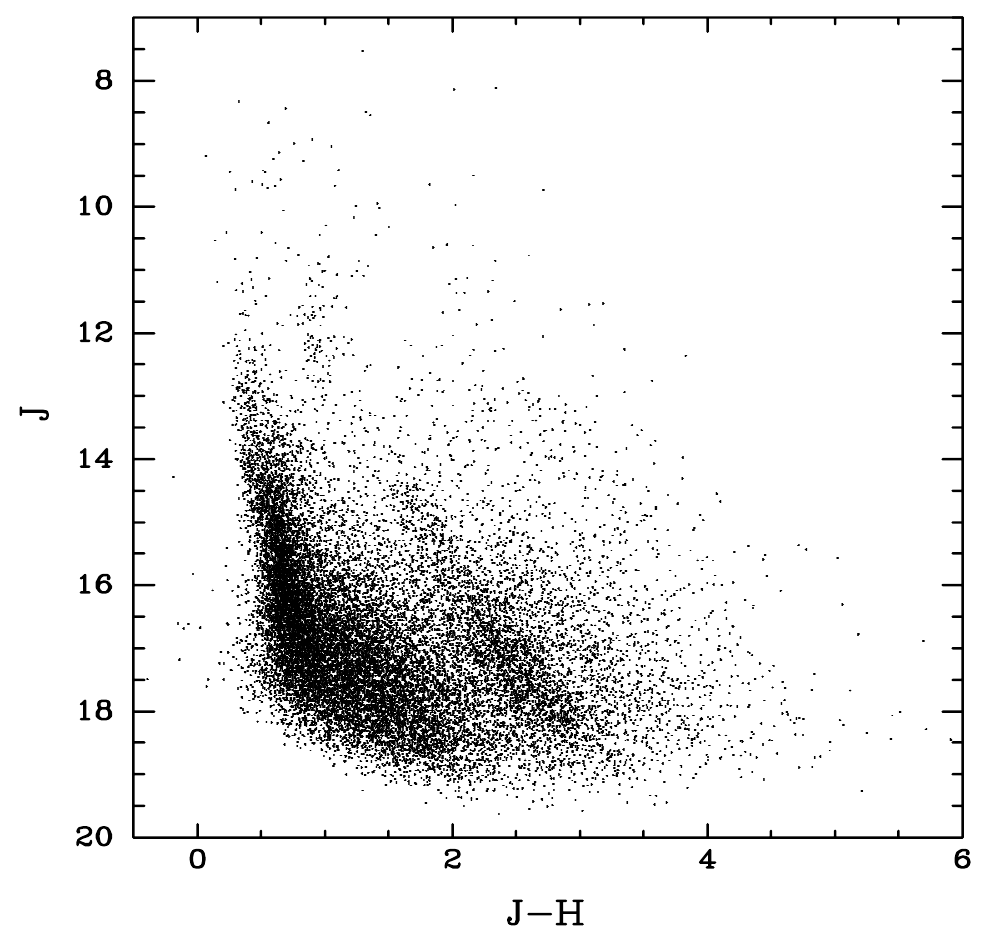}
\caption []{Near-infrared $J-H,J$ diagram of all the stars detected in the field. Different populations can be identified and are discussed in the text.}
\label{JH_J}
\end{center}
\end{figure}

\begin{figure}[ht]
\begin{center}
\hspace{-0.5cm}
\includegraphics [width=8.5cm, angle={0}]{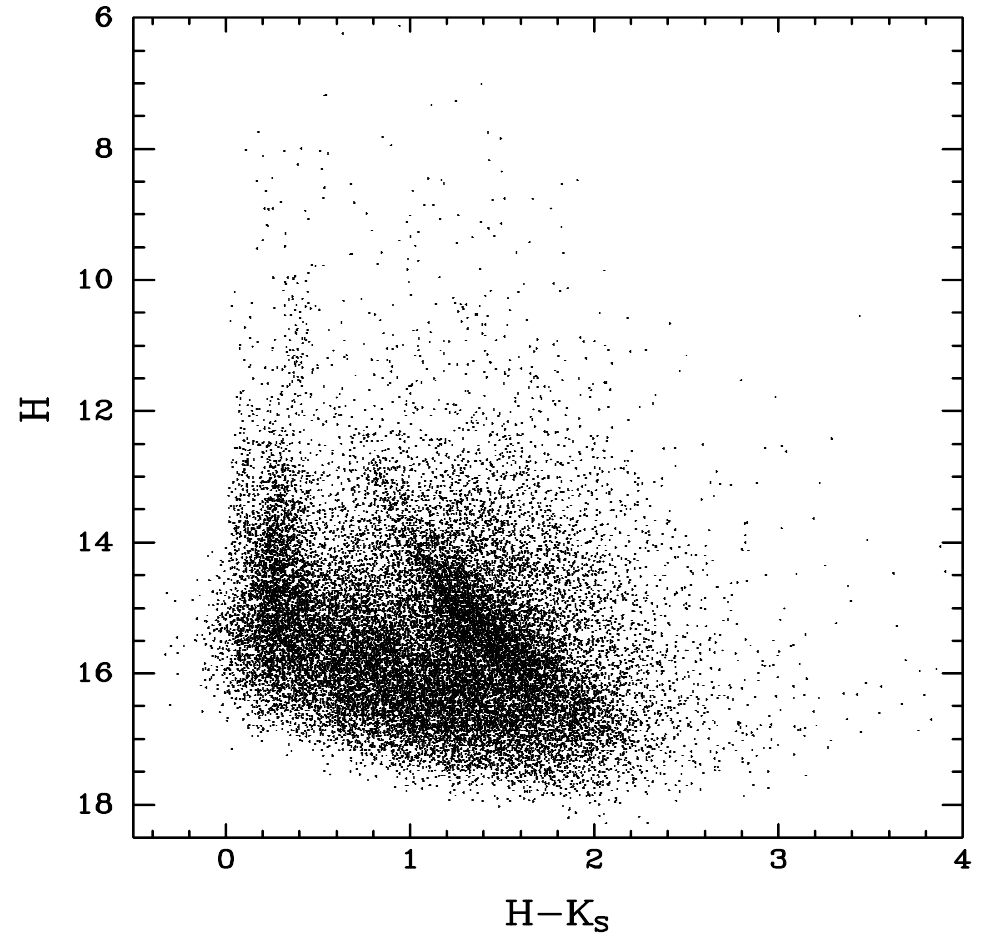}
\caption []{Same as Figure~\ref{JH_J}, now for the $H-Ks, H$ diagram of all the stars detected in the field.}
\label{HKs_H}
\end{center}
\end{figure}

\subsection{Color-magnitude diagrams\label{colmag}}

The color-magnitude diagrams of the stars in the field, presented in Figs.~\ref{JH_J} and \ref{HKs_H}, are typical of
a low galactic latitude field in the direction to the inner regions of the Milky Way. \citep[e.g.,][]{Gonzalez12}. The almost vertical locus with low color indices defined by local, unobscured main sequence stars can be traced especially well in the $(H-K_S, H)$ diagram, well separated by $\Delta (H-K_S) \simeq 0.3$ from the sequence formed by local giants and subgiants. A most prominent feature in both the $(J-H, J)$ and the $(H-K_S, H)$ diagrams is the band starting at $J-H \simeq 1.6$, $J \simeq 14.5$ and running diagonally toward redder colors and fainter magnitudes along the reddening vector, marking the location of red giant branch stars in the galactic bulge.

Both color-magnitude diagrams also show a distinct group of bright red stars with $11.0 < J < 12.5$, $J-H \simeq 0.9$ (Fig.~\ref{JH_J}) or $10.0 < H < 11.5$, $H-K_S \simeq 0.4$ (Fig.~\ref{HKs_H}). This is reminiscent of a feature first identified by \citet{Hammersley00} and confirmed by subsequent works \citep{Lopez07}, which is interpreted as the signature of the galactic bar traced by the red clump, consisting of cool, high-metallicity horizontal branch stars burning helium in their cores \citep{Seidel87,Paczynski98}. A detailed examination shows however that this feature in our color-magnitude diagrams is not related to the galactic bar. The group in our results is on the average brighter by $\simeq 1.0$~magnitudes at the same extinction\footnote{The extinction law of \citet{Cardelli89} with a total-to-selective extinction ratio $A_V / E[B-V] = 3.1$ is used throughout this study; specifically, we use $A_J/A_V = 0.282$, $A_H/A_V = 0.190$, $A_{Ks}/A_V = 0.114$.} than the red clump stars at the near end of the bar at $l = 27^\circ$, whereas we are looking in the direction of $l = 16^\circ 36$ where the bar is more distant. Adopting an inclination angle of the bar of $45^\circ$ \citep{Lopez07} places the bar in the direction $l = 16^\circ 36$ at a distance 8\% greater than the near end of the bar, thus making the red clump stars tracing it fainter by 0.17~mag. Adopting a distance to the galactic center of 8.0~kpc, this places the red clump at 6.4~pc from the Sun in the direction $l = 16^\circ 36$ while the relatively blue colors of our group of stars requires an extinction below $A_V \sim 3$~mag, which is hardly compatible with such a large distance so close to the galactic midplane direction, and is furthermore in conflict with other characteristics of the extinction in our field as discussed in Sections~\ref{evidence} and \ref{results}. We identify the group of bright, lightly reddened stars discussed here as the local population of red giant stars. Some implications of the fact that such stars in our field reach down to $J = 12.5$, $H = 11.5$ with a relatively low amount of foreground reddening will be discussed in Sect.~\ref{discussion}.

\subsection{Color-color diagram and the evidence for an obscured population of hot stars\label{evidence}}

\begin{figure}[ht]
\begin{center}
\hspace{-0.5cm}
\includegraphics [width=8.5cm, angle={0}]{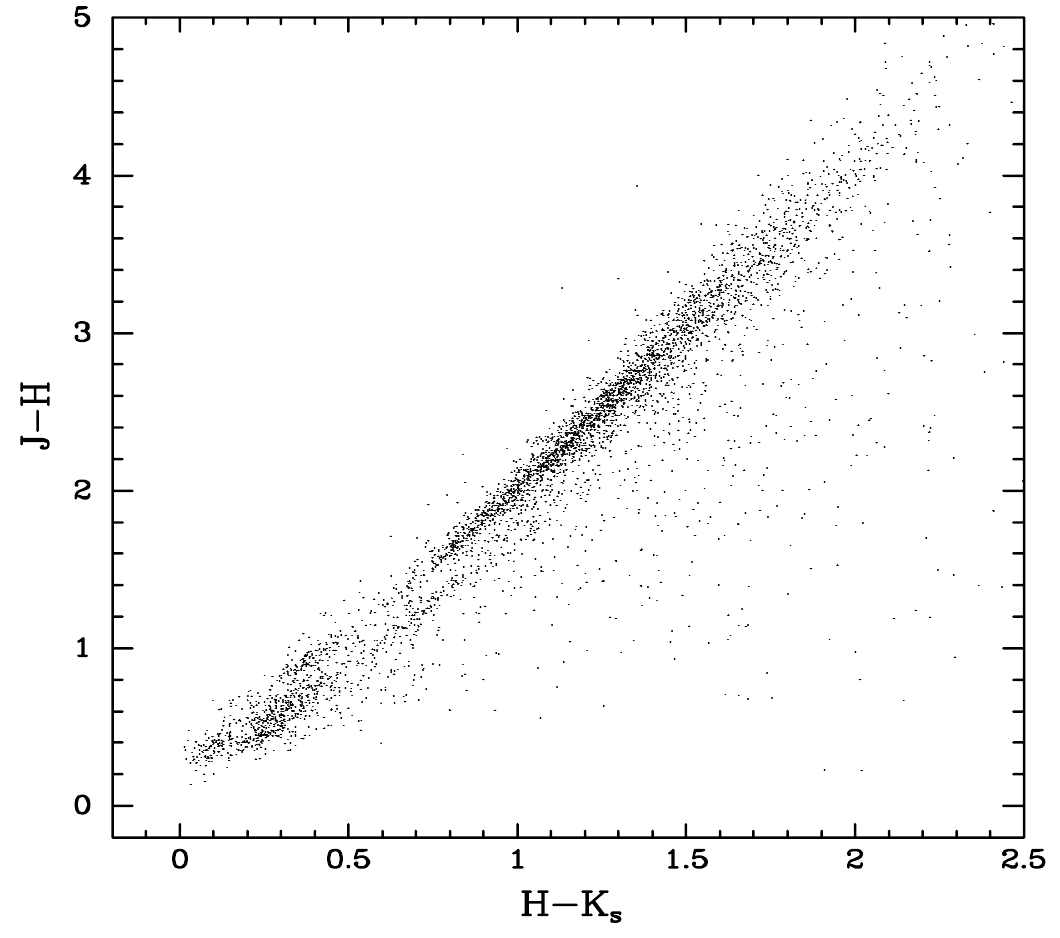}
\caption []{Color-color diagram of the imaged field. For clarity, only stars with $K_S < 13.5$ are plotted. The two parallel sequences along the reddening vector, respectively associated with cool (upper band) and hot (lower band) stars,are clearly seen. The gaps along those bands are interpreted as features caused by extinction.}
\label{HKs_JH}
\end{center}
\end{figure}

The $(H-K_S)$, $(J-H)$ color-color diagram of the stars in the field is presented in Figure~\ref{HKs_JH}. To avoid
contamination and loss of the detailed structure of the diagram by the lower quality of the photometry of faint sources, the diagram is restricted to stars brighter than $K_S = 13.5$. To a first approach the diagram is typical of a field in a direction near the galactic plane toward the inner regions of the Galaxy, with sources covering a wide range in foreground extinction that results in a wide range of values in each color and a long sequence of points along the reddening line.

A close look shows some interesting features. Most of the stars with lightly reddened colors (up to $H-K_S \simeq
0.5$) cluster toward the lower end of the reddening sequence, as expected from a magnitude-limited field population dominated by main sequence stars that includes nearby cool dwarfs with relatively red $(H-K_S)$ colors at a given value of $(J-H)$. The bright, lightly reddened local giants discussed in the previous Section are also clearly distinguished in the diagram as the concentration of points centered near $(H-K_S) = 0.4$, $(J-H) = 1.0$. Their location on the extension to low reddening of the main band of points in the diagram, marking the location of distant, reddened cool giants, confirms that these sources are also cool giants and not reddened main sequence stars.

The rather sharp start of the main band of stars along the reddening vector at $(H-K_S) = 0.75$, $(J-H)=1.6$ is caused by the beginning of the dominance of the bulge and it is the counterpart of the corresponding band of bulge giants already discussed in Sect.~\ref{colmag}. However, this alone does not explain the decline in the population of that band in the $0.5 < (H-K_S) < 0.75$ interval as compared with its higher crowding at bluer colors. We attribute the decline to the existence of a jump in extinction causing a drop in the counts of cool giants with observed colors $(H-K_S) > 0.5$, corresponding to a reddening of $E(H-K_S) > 0.38$ ($A_V = 5.0$) when taking $(H-K)_0 = 0.12$ as the intrinsic color of a K2III star \citep{Tokunaga00}, the most abundant type of red giants.

The most relevant feature of the color-color diagram for the present study is the obvious split of the band populated
by reddened stars into two parallel branches, most obvious in the $0.6 < (H-K_S) < 1.0$ interval. The lower branch may in principle be formed by either late-type dwarfs or by very early-type stars with slightly negative $(H-K_S)_0$, $(J-H)_0$ intrinsic colors, both significantly reddened. The former option is discarded due to the combination of the intrinsic faintness of late-type dwarfs and the large distance implied by the foreground extinction, leaving hot stars as the only population able to produce the trace. The population of this parallel reddening band is the hallmark of presence of OB stars and has been used to reveal many new obscured members with spectral types B2 and earlier of OB associations in Cygnus \citep{Comeron02,Comeron08,Comeron12}. The prominence of this band in the field imaged by us demonstrates that a substantial population of moderately reddened OB stars does exist in that direction.

\section{Discussion\label{discussion}}

Are the newly identified early-type stars members of a massive population around RAFGL~5475 responsible for the shell-like structures and the ionized regions detected at long wavelengths? The high obscuration in their direction at visible wavelengths, together with their moderately faint magnitudes (the brightest member of this group has $J = 11.2$) places them beyond the capabilities of Gaia to provide a direct answer to the question by measuring their trigonometric parallaxes. Radial velocity measurements through intermediate resolution infrared spectroscopy are a more promising venue, as they may show velocities close to those of the interstellar gas structures noted in Section~\ref{intro}, but the imaging results available already support their likely membership.

As shown in Fig.~\ref{HKs_JH} and discussed in Sect.~\ref{evidence}, the band of lightly reddened cool giants stops
rather abruptly at the colors corresponding to an extinction $A_V = 5.0$, and then resumes at a $(H-K_S) > 0.8$, implying an extinction $A_V = 9.0$ or greater for the intrinsic colors of a typical K2III giant. This leads us to infer the existence of a "wall" of dust rising the extinction from $A_V \simeq 5.0$ to $A_V \simeq 9.0$, located so that the group of local giants noted in Figs.~\ref{JH_J} and \ref{HKs_H} are in front of it. The members of the group of lightly reddened cool giants are found down to $J \simeq 12.5$, $H \simeq 11.5$ which, at the limiting extinction of $A_V = 5.0$ derived above indicates a distance modulus $DM < 12.0$ (thus a distance of up to 2.5~kpc) for this group adopting $M_J = -0.9$ as the absolute magnitude of K2III giants. This distance suggests that the extinction may be caused by foreground clouds at the distance of M16 and M17, and particularly with the half-shell seen both in HI and CO that includes both regions \citep{Moriguchi02}. Adopting $(H-K_0) = -0.05$ as the intrinsic color of the reddened early-type stars noted in Section~\ref{evidence}, the beginning of the band that they occupy at $(H-K_S) = 0.6$ also implies $A_V \gtrsim 9$~mag, thus implying that they are located behind that extinction wall.

The distribution of the CO emission in the general direction of the field around RAFGL~5475 has a local peak in the interval $+40 < v_{\rm LSR} {\rm (km \ s^{-1})} < +50$, where the HII region excited by RAFGL~5475 and the other interstellar features of the region appear, and is even stronger in the interval $+45 < v_{\rm LSR} {\rm (km \ s^{-1})} < +50$, where the structures directly associated with RAFGL~5475 lie. The dark cloud surrounding RAFGL~5475, whose silhouette is obvious near the center of Figure~\ref{JHKs}, is most likely the cause of the peak in the latter range. In the CO spectrum, the peak is superimposed on the slope of a broader emission feature that extends from $+35 < v_{\rm LSR} {\rm (km \ s^{-1})} < +52$ in velocity and continues toward higher galactic longitudes, well beyond the boundaries of our field, probably tracing the large-scale distribution of molecular gas in the Scutum-Centaurus arm. By removing a baseline obtained by interpolating the spectrum in the direction $l = 16^\circ 375$, $b = -0^\circ 25$ between velocities outside the interval $+45 < v_{\rm LSR} {\rm (km \ s^{-1})} < +50$, the residual spectrum of the peak associated with RAFGL~5475 has a CO equivalent width $W_{\rm CO} = (17.6 \pm 5.0)$~K~km~s$^{-1}$, which translates into $N_H = (3.2 \pm 0.6) \times 10^{21}$~cm$^{-2}$ using the conversion $N_H = 1.8 \times 10^{20} W_{\rm CO}$~cm$^{-2}$~K~km$^{-1}$~s \citep{Dame01}. That column density integrated over the area traced by bright nebulosity in Spitzer images yields an estimated mass of the cloud of 2000 -� 3000 M$\odot$.

We identify candidate members of the association by defining the reddening-free parameter $Q = 0.019 - 0.443 (J-H) + 0.891 (H-K_S)$ that measures the distance of a point on the $(H-K_S)$, $(J-H)$ diagram to the center of the band traced by the red giants. Early-type stars are found at positive values of $Q$ and their sequence in Figure~\ref{HKs_JH} is defined by a narrow range of values, $0.07 < Q < 0.13$\footnote{It may be noted that the coefficients of the definition of this parameter differs slightly from those adopted in previous works like \citet{Comeron02,Comeron08} or \citet{Comeron12}, but the principle remains the same}. This locus has some degree of contamination by unrelated objects as shown by its application to other regions where the classification of the early-type candidates could be spectroscopically verified \citep{Comeron12}, but the fact that it is clearly defined in Figure~\ref{HKs_JH} indicates that most of the objects populating it are indeed early-type stars. Conversely, early-type stars may have slightly deviant colors placing them outside the proposed range of $Q$ values given above. Our list of candidates is therefore neither uncontaminated nor complete, but it can still be used as a first approximation to study the stellar content and distribution of the massive members of the RAFGL~5475 association.

\begin{figure}[ht]
\begin{center}
\hspace{-0.5cm}
\includegraphics [width=8.5cm, angle={0}]{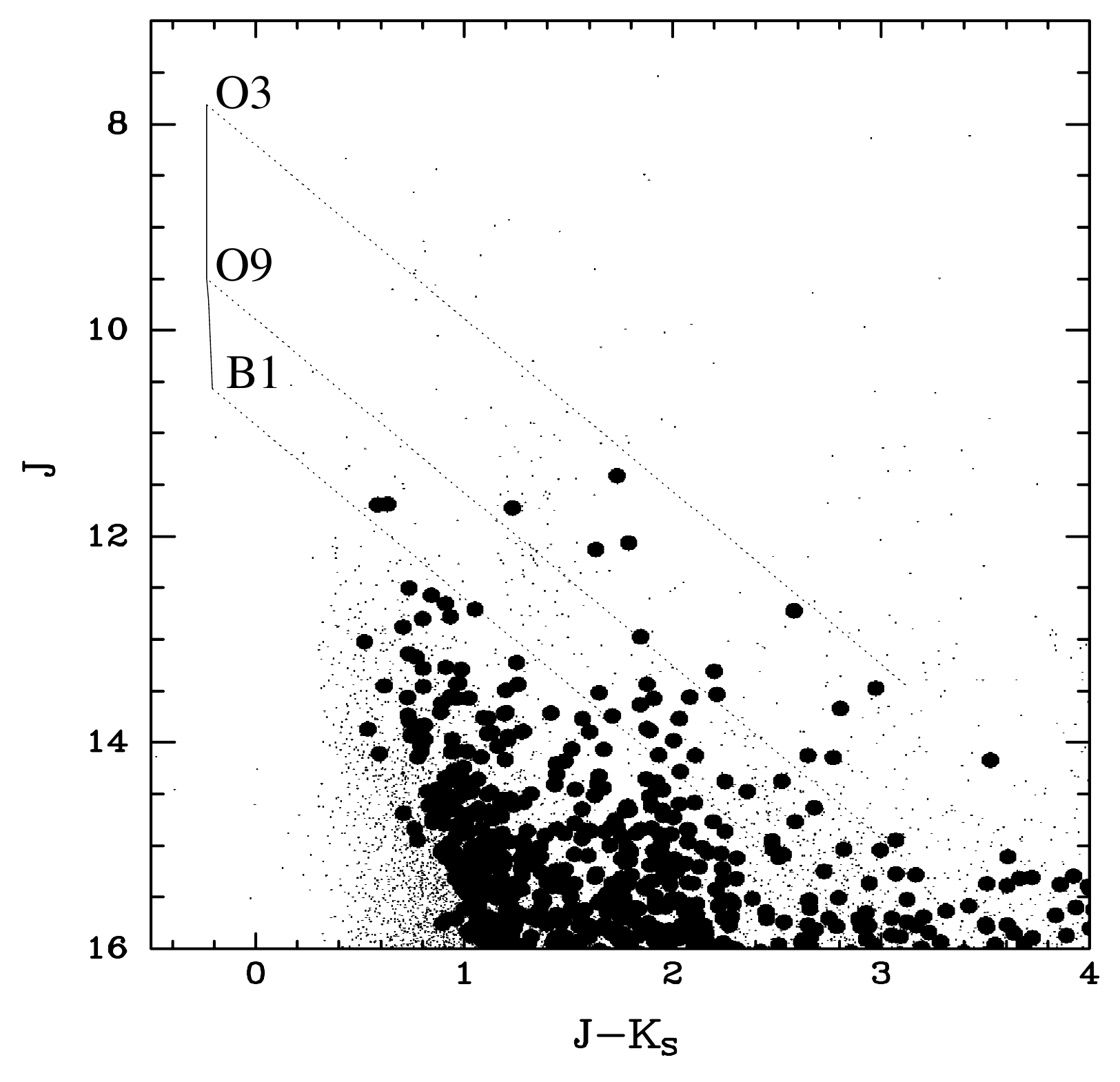}
\caption []{$J-K_S, J$ diagram of all the stars in the imaged field. Filled circles correspond to early-type candidates selected through the $Q$ criterion described in Section~\ref{discussion}. The location of very early-type stars of different spectral types at the distance of RAFGL~5475 obscured by varying amounts of foreground extinction is indicated with the dotted lines, and their unreddened magnitudes at that distance are given as the solid line near the top left edge of the diagram. The corresponding lines for a hypothetical population at the distance of M16 and M17 would be shifted upwards by 1.5 magnitudes. The total length of each line corresponds to a visual extinction $A_V = 20$~mag. }
\label{JKs_J_withOB}
\end{center}
\end{figure}

\begin{table*}[t]
\caption{Candidate B1 and earlier-type members of the RAFGL 5475 association}
\begin{tabular}{cccccc}
\hline
RA(2000)     & Dec(2000)   &         $J$      &    $J-H$        &    $H-K_S$      \\
\hline
 18:21:32.05 & -14:46:55.9 & 11.412$\pm$0.005 & 1.095$\pm$0.008 & 0.639$\pm$0.007 \\
 18:21:10.13 & -14:41:07.2 & 12.064$\pm$0.002 & 1.147$\pm$0.004 & 0.642$\pm$0.005 \\
 18:21:29.99 & -14:42:01.0 & 12.127$\pm$0.004 & 1.048$\pm$0.005 & 0.584$\pm$0.007 \\
 18:20:45.11 & -14:43:52.9 & 12.976$\pm$0.008 & 1.151$\pm$0.012 & 0.696$\pm$0.013 \\
 18:21:17.58 & -14:48:10.5 & 13.309$\pm$0.010 & 1.393$\pm$0.012 & 0.807$\pm$0.008 \\
 18:21:29.90 & -14:52:04.0 & 13.434$\pm$0.010 & 1.207$\pm$0.013 & 0.670$\pm$0.012 \\
 18:21:05.52 & -14:38:46.8 & 13.517$\pm$0.003 & 1.022$\pm$0.007 & 0.626$\pm$0.008 \\
 18:21:33.64 & -14:48:41.0 & 13.534$\pm$0.007 & 1.422$\pm$0.010 & 0.791$\pm$0.009 \\
 18:21:44.68 & -14:47:55.5 & 13.559$\pm$0.007 & 1.319$\pm$0.009 & 0.764$\pm$0.008 \\
 18:21:17.15 & -14:40:06.2 & 13.573$\pm$0.003 & 1.213$\pm$0.007 & 0.696$\pm$0.008 \\
 18:20:54.64 & -14:38:43.8 & 13.634$\pm$0.021 & 1.159$\pm$0.042 & 0.686$\pm$0.052 \\
 18:21:39.97 & -14:49:40.1 & 13.743$\pm$0.010 & 1.074$\pm$0.014 & 0.637$\pm$0.011 \\
 18:21:25.61 & -14:42:39.2 & 13.769$\pm$0.006 & 1.299$\pm$0.007 & 0.734$\pm$0.008 \\
 18:21:16.15 & -14:55:36.2 & 13.864$\pm$0.012 & 1.183$\pm$0.017 & 0.694$\pm$0.016 \\
 18:21:26.78 & -14:45:46.6 & 13.885$\pm$0.007 & 1.188$\pm$0.008 & 0.703$\pm$0.006 \\
 18:20:58.50 & -14:43:55.7 & 13.981$\pm$0.005 & 1.260$\pm$0.008 & 0.745$\pm$0.008 \\
 18:21:16.77 & -14:54:38.0 & 14.122$\pm$0.012 & 1.227$\pm$0.017 & 0.706$\pm$0.020 \\
 18:20:56.90 & -14:41:23.6 & 14.126$\pm$0.005 & 1.333$\pm$0.009 & 0.776$\pm$0.009 \\
 18:21:24.88 & -14:41:10.3 & 14.282$\pm$0.008 & 1.298$\pm$0.010 & 0.739$\pm$0.011 \\
 18:21:54.67 & -14:38:51.2 & 14.378$\pm$0.008 & 1.432$\pm$0.010 & 0.819$\pm$0.011 \\
 18:20:54.09 & -14:49:21.0 & 14.475$\pm$0.024 & 1.517$\pm$0.027 & 0.841$\pm$0.015 \\
\hline
\end{tabular}
\label{members}
\end{table*}

Fig.~\ref{JKs_J_withOB} shows a color-magnitude diagram of the region with all the stars with $Q$ in the range given above highlighted. Also shown is the location of unreddened main sequence O and early B stars at a distance of 4.0~kpc, together with the reddening vectors starting at selected spectral types. Intrinsic colors and magnitudes are taken from the compilation of \citet{Pecaut13} \footnote{We have used the expanded online version of the table of intrinsic properties of main sequence stars that can be found at {\tt http://www.pas.rochester.edu/$\sim$ emamajek/EEM$\_$dwarf$\_$UBVIJHK$\_$colors$\_$Teff.txt}, which includes an absolute magnitude vs. spectral type calibration.}. The increasing crowding of points at the faintest $J$ magnitudes is largely due to the influence of photometric uncertainties, which spuriously populate the $0.07 < Q < 0.13$ band. The same effect happens at the highest reddenings, where small photometric errors of even a small percentage of the much more abundant distant cool giants can dramatically increase the contamination of the early-type locus. However, as hinted by Fig.~\ref{HKs_JH} the area of the diagram populated by bright stars with moderate reddening stands out as dominated by a genuine early-type population, and the diagram in Fig.~\ref{JKs_J_withOB} suggests the presence of several O and B stars in the region. We note that Fig.~\ref{JKs_J_withOB} rules out the presence in our field of any lightly reddened foreground OB stars that might be associated with the M16/M17 region: all the lightly reddened stars ($J-K_S \lesssim 1.5$) fulfilling the $Q$ selection criteria are much fainter than the unreddened early-type main sequence locus at the distance of M16/M17, which is displaced by 1.5~mag toward brighter magnitudes with respect to the one depicted in Figure~\ref{JKs_J_withOB}.

\begin{figure}[ht]
\begin{center}
\hspace{-0.5cm}
\includegraphics [width=9cm, angle={0}]{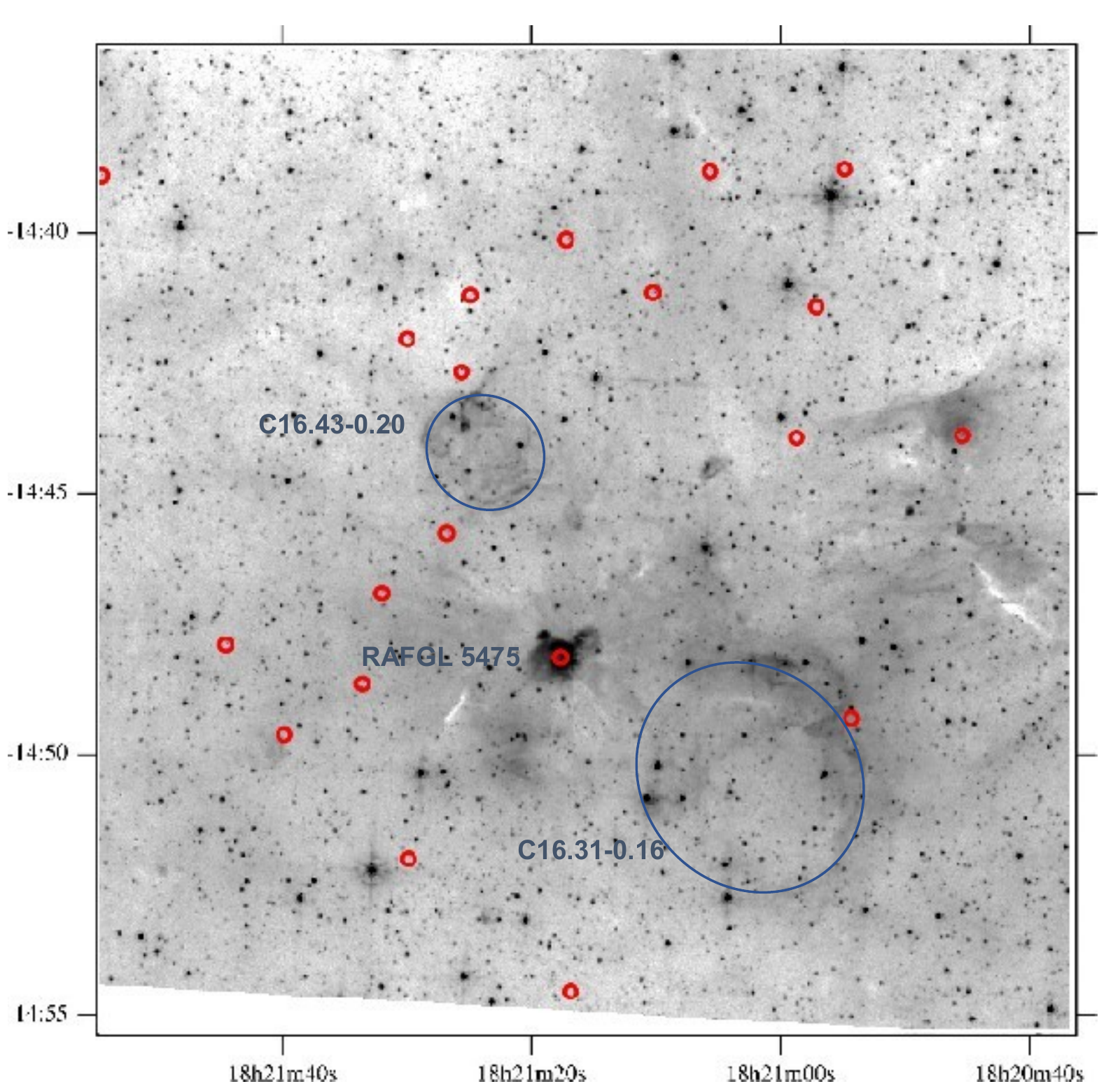}
\caption []{Positions of the candidate stars earlier than B1 listed in Table~\ref{members}, superimposed on an archive {\sl Spitzer} image at 5.8~$\mu$m. The southernmost star in Table~\ref{members}, at RA(2000)~=~18:21:16.15, Dec(2000)=-14:55:36.2, is just outside the boundary of the field presented here in a zone that was not covered by {\sl Spitzer}, and it is not plotted in the figure.}
\label{overlay}
\end{center}
\end{figure}

Figure~\ref{overlay} displays the location of all the candidate early-type stars with $(J-K_S)< 2.4$ that lie above the reddening vector having its origin at the unreddened B1 main sequence type in Fig.~\ref{JKs_J_withOB},that is, a sample of bona fide OB star candidates, overlaid on a Spitzer image of the field at 5.8~$\mu$m showing a wealth of structures in the interstellar medium that are undetected at the shorter wavelengths of our observations. We restrict the list of candidates to such bright early-type stars to avoid contamination by field stars or by giants erroneously moved to the {\sl locus} of early-type stars by errors in the photometry. Our candidate OB stars, which are listed in Table~\ref{members}, are well spread over the field generally lying projected on the areas of infrared nebulosity. There are no apparent signs of clustering or a concentration toward to the most conspicuous interstellar medium features, with the exceptions of the star at RA(2000)~=~18:20:45.11, Dec(2000)~=~-14:43:52.9 projected on a bright knot of isolated nebulosity; the star at RA(2000)~=~18:20:54.09, Dec(2000)~=~-14:49:21.0, which appears projected on a brightening of the nebula associated with the HI shell C16.31-0.16 \citep{Anderson09} and most notably the star at RA(2000)~=~18:21:17.58, Dec(2000)~=~-14:48:10.5, which is projected only 10'' from RAFGL~5475 itself (0.2~pc projected distance), although far less embedded. Remarkably, no OB star appears within the well delineated shell C16.43-0.20.

\section{Summary and conclusions}\label{conclusions}

In this study we have investigated the possible presence of very early-type stars in the direction of RAFGL~5475, a bright infrared source virtually unexplored thus far that is surrounded by structures in the interstellar medium usually resulting from the energetic action of massive stars. We have identified a number of early-type star candidates obscured by a significant amount of foreground extinction. The high extinction, together with the evidence for much of its origin in an obscuring layer at the same radial velocity of RAFGL~5475 and its associated structures, leads us recognize these stars as members of a new OB association lying 4~kpc from the Sun in the Scutum-Centaurus spiral arm of our Galaxy. The association seems to host a number of O-type stars, some of them positionally coincident with distinct structures in the interstellar medium. RAFGL~5475 is most likely a newly born massive star, still heavily embedded in its parental cloud, perhaps representing the latest stage of massive star formation in the association. The $\sim 25$~pc extent of the association, its stellar content, the mass of the main associated molecular cloud estimated to be $2000 - 3000$~M$_\odot$, and the evidence for massive star formation spread over time are reminiscent of star formation in the Orion Molecular Cloud.

The limited observational material at hand calls for further characterization of the newly discovered RAFGL~5475 association. New observations should lead to the confirmation and accurate classification of the candidate members identified thus far, to a more complete and robust census of members of the association, to the possibility of searching for clustering and structure in the distribution of the members, and to the investigation of their relationship with the structures observed in the interstellar medium.

\begin{acknowledgements}

It is always a pleasure to acknowledge the excellent support provided by the staff at the Calar Alto Observatory,
especially on this occasion by Gilles Bergond, Ana Guijarro, and David Galad\'\i . The comments by the referee of the first version of the manuscript were of great help in guiding some significant improvements. The Two Micron All Sky Survey (2MASS) is a joint project of the University of Massachusetts and the Infrared Processing and Analysis Center/California Institute of Technology, funded by the National Aeronautics and Space Administration and the National Science Foundation. This research has made use of the SIMBAD database, operated at CDS, Strasbourg, France.

\end{acknowledgements}

\bibliographystyle{aa} 
\bibliography{rafgl5475_cit}





\end{document}